# Triggering of Aftershocks of the Japan 2011 Earthquake by Earth Tides

A. Datta and Kamal*

Department of Earth Sciences, Indian Institute of Technology Roorkee, Roorkee-247667, Uttarakhand, India
email: arjun_datta23@hotmail.com ; kamalfes@iitr.ernet.in*
* Corresponding author

**The aftershock sequence of the devastating Japan earthquake of March 2011 is analyzed for the presence of periodicities at the Earth tide periods. We use spectral analysis as well as a time-domain method KORRECT developed earlier to detect presence of diurnal and semi-diurnal periodicities in the sequence of aftershocks (M ≥ 4). This suggests that large aftershocks in the fault zone of the Japan 2011 earthquake were strongly influenced by Earth tides.**

The elastic rebound theory is now well accepted as the reason for the occurrence of earthquakes. When fault stresses rise above a critical threshold for rupture[1], the Earth slips along the fault and an earthquake results. It is then only intuitive that any additional stress acting on a fault system that approaches failure could trigger the rupture process that produces the earthquake. The stresses responsible for the occurrence of an earthquake are tectonic in origin but the final onset of rupture can be affected by other kinds of stresses superimposed on tectonic stresses during the build-up to failure. We use the term 'affected' here because the additional stress may advance or retard the onset of rupture, depending on the sense in which it acts.

Perhaps the strongest candidate for earthquake triggering is the Earth tide. Gravitational attractions, primarily of the sun and moon, cause periodic elastic deformation of the solid Earth, thus exerting additional stresses. These tidal stresses are oscillatory in nature and their magnitude is of the order of $10^3$ Pa. This is much less than average stress drops in earthquakes[2] ($10^5$-$10^7$ Pa) but the rate of tidal stress change is much higher than the build-up rate of tectonic stress[3] rate and this makes tidal triggering of earthquakes possible.

The question of whether earthquakes are triggered by Earth tides was raised over a century ago[4, 5] and since then, numerous efforts have been directed towards answering it. These studies investigated different kinds of data-sets and employed a variety of methods[6], but owing to mixed results[2, 7-10], no conclusive evidence had emerged until recently for earthquake triggering by the Earth tide. In the last decade, tests on tidal triggering have produced more homogeneous and positive results. A breakthrough was achieved very recently[11] when the triggering effect of the earth tide was demonstrated in earthquakes of all magnitudes (>2.5) and all types of focal mechanisms, using the largest global earthquake catalog (NEIC world seismic catalog) available. In 2004, a study of seismicity in Japan[12] found that regions which experienced a large earthquake showed the best earthquake-tide correlations. These two findings were the motivation of our search for tidal triggering of aftershocks of the recent earthquake in Japan. Additionally, an aftershock sequence is a type of data-set that is amenable to the simple methods of analysis employed in this study[13].

The Tohoku-Oki earthquake (magnitude, Mw=9) off the East coast of Honshu, Japan, occurred at 05:46:24.56 UT on 11$^{th}$ March 2011. Epicentre location was $38.299^0$ N, $142.373^0$ E and focal depth was 32km. At the time of this investigation, an almost complete record of aftershocks of magnitude $M \geq 4$ occurring in the first 30 days following the mainshock was available from the NEIC PDE catalog of the US Geological Survey. Aftershock data obtained consisted of origin time, magnitude, epicentre location and focal depth. Since we are interested only in those events which occurred in the fault zone of the mainshock, we have restricted this investigation to those events which occurred in the rectangular area between latitudes $34^0$ to $41^0$ N and longitudes $139.5^0$ to $145.5^0$ E. In this region, 1,370 aftershocks (of $M \geq 4$) have been recorded by the USGS in the one month following the mainshock. These numbers make our data-set much larger than those used in previous studies of triggering of aftershocks[14, 15]. We tested the completeness of the catalog by linear least-squares fitting of the Guttenberg-Richter relation to the data and found a b-value of ~1.04 which is taken as acceptable. Figure 1 shows the magnitude and depth distributions of aftershocks.

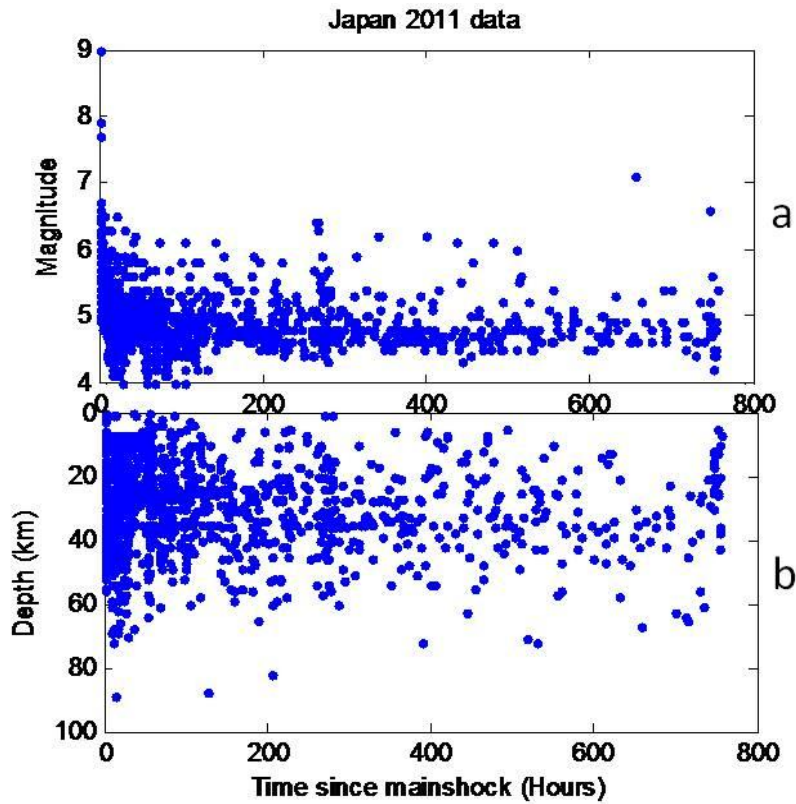

**Figure 1**: a) Magnitude and b) Depth distributions of the entire Japan aftershock data-set during first 30 days after the mainshock

In this study, we first subject the aftershock sequence to spectral analysis. Before this is done however, the sequence needs to be modified in order to obtain an equispaced time series with some meaningful amplitude. The time axis is divided into fixed width bins and the number of events in each bin is counted. This number is ascribed to the centre of the bin. The result is a time series with sampling interval equal to the bin-width and amplitudes equal to the number of events per bin. Different bin-widths, ranging from 20 min to 1 hr were tried. All of these would allow the detection of frequencies several times higher than the strong tidal components. Each binned time series was tapered with a Tukey window and transformed to the frequency domain with the FFT routine in MATLAB.

The spectrum was computed for each of the binned series for the first 240 hours (10 days) of data. Figure 2 shows the power spectra for the two different values of bin width. The two distinct peaks are at 1 cycle per day and 1.9 cycles per day, corresponding to the diurnal and semi-diurnal components of the Earth tide. The same peaks were obtained in the spectra of all the binned series.

The 240-hr window was then shifted forward in time and the spectrum was computed for each window position. It was found that by the end of the 12<sup>th</sup> day after the mainshock, it is impossible to detect any periodicities in the spectra. This is shown as a part of Figure 4 and is attributed to the sharp decline in the number of aftershocks with time. Our faith in the presence of tidal peaks in the power spectra was strengthened when we performed an identical analysis on the aftershock data sets from three other large earthquakes for which similar-size data was available from the same source – Indonesia 2004 (offshore Northern Sumatra, Mw 9.1), Indonesia 2005 (Northern Sumatra, Mw 8.6) and Chile 2010 (offshore Bio-Bio, Mw 8.8). These had 1,361, 1,253 and 1,353 aftershocks in 30 days respectively. Figure 3 shows the spectra obtained for a particular bin width; similar results were obtained for other bin widths. Power spectra of series constructed from these data sets do not show any evidence of presence of tidal periodicities and no account of tidal triggering of aftershocks of these earthquakes has ever been reported.

Encouraged by the results of the spectral analysis, we employed a second, time-domain method known as KORRECT[13] to substantiate our results.

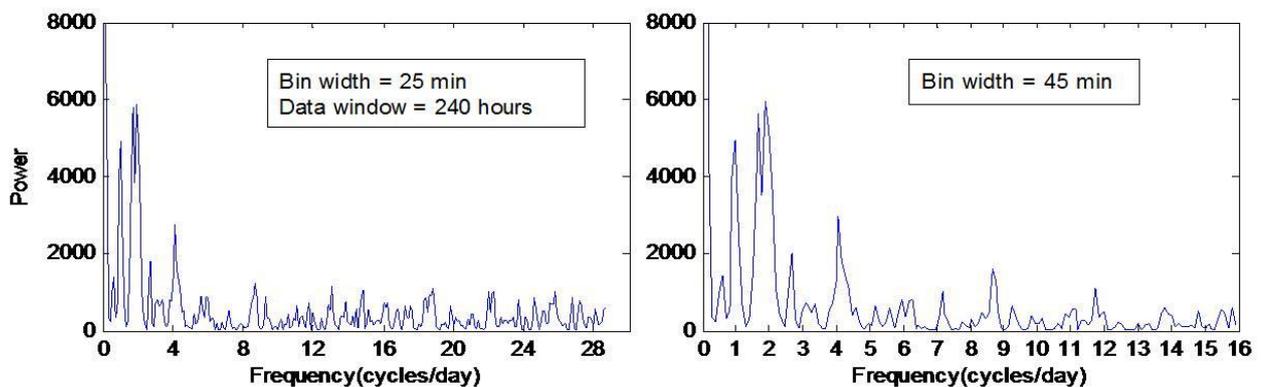

**Figure 2**: Power spectrum for two different binned series. The time series begins at the mainshock and ends at 10 days after the mainshock. In both cases, the peaks appear exactly at the same positions (1 and 1.9 cycles per day). Note the consistent nature of the peaks.

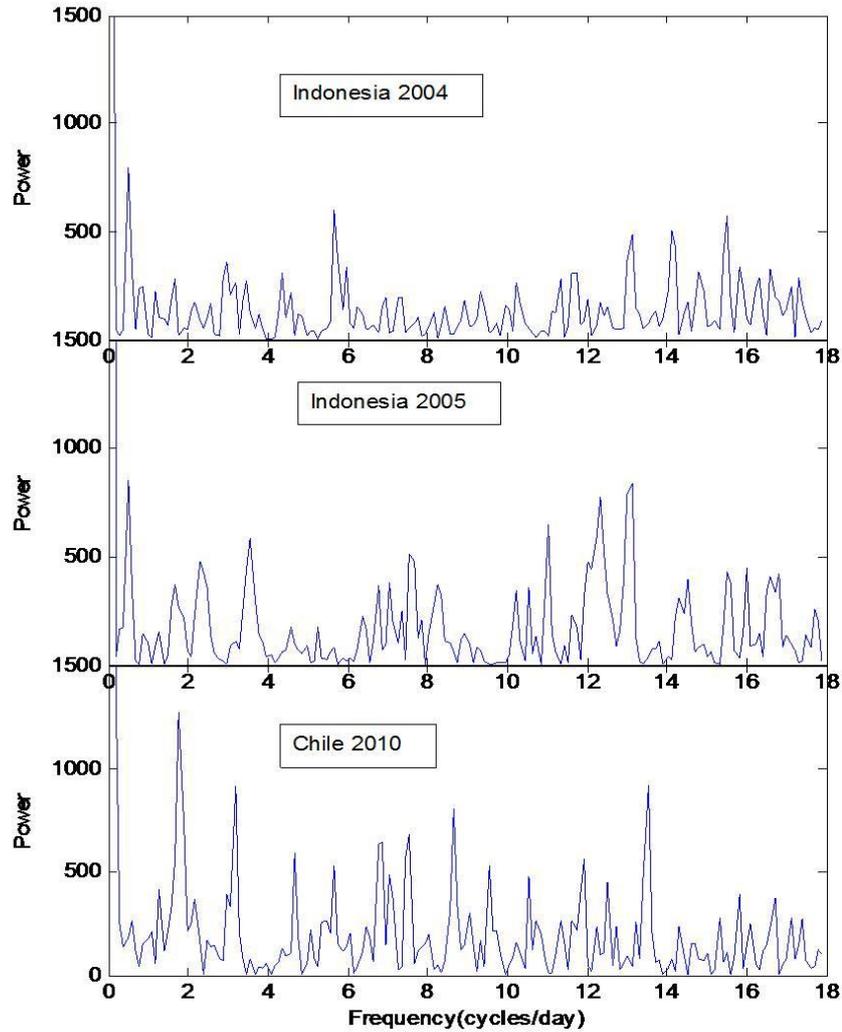

**Figure 3**: Power spectrum of aftershock sequences of three other M>8 earthquakes. In each case the bin width was 40 min and the data used was from the first 240 hours following the mainshock. There are no peaks at tidal periods that rise above the background noise level in the signal, implying that tidal periodicities are not present in these sequences.

In this method, a pulse series is formed by repeating a finite-width rectangular pulse at regular intervals. The pulse-width used is the bin-width used in the first method and the pulse separation, or the period of the pulse series, is made equal to the harmonic periods corresponding to the frequency samples obtained in the spectral analysis. This gives $N/2$ different periods for the pulse series, where $N$ is the number of time samples used to compute the FFT. For every such period, the pulse series is overlaid on the aftershock sequence in its original form. All the events falling within the pulses are counted and added up. This gives the number of aftershocks which have occurred at a particular phase of the period. On dividing this number by the total number of pulses that span the duration $T$ of aftershock data, we get a measure of the aftershock 'density' (events per cycle) at a particular phase of the period. The pulse series is then shifted along the time axis, with pulse separation kept constant, to compute the aftershock density at another phase of the same period. We performed phase shifts in steps of $5^0$. A plot of the aftershock density as a function of phase gives the 'phase variation curve' for a particular period. If the phase variation curve for a particular period shows a significant rise/fall in aftershock density for certain phases, that indicates that in every cycle of an oscillating sequence with that period, anomalously high/low number of aftershocks during that portion of the cycle. This clearly suggests triggering of aftershocks by that oscillating sequence. Most of the phase variation curves obtained in this analysis were not of this type. They were either flat or exhibited oscillatory behaviour around a mean value. However, when the period of the pulse-series was made equal to the period corresponding to the frequencies at which peaks appear in the power spectra (periods of 12.63 hours and 24 hours), we obtained phase variation curves with a distinct character.

This can be seen in parts a) and b) of Figure 4 for the 12.63 hr period. Power spectra have also been shown in the left panel for comparison. It is clear from Figures 4 a) & b) that more aftershocks occur during a certain portion of the semi-diurnal cycle, implying that certain phases of the cycle are favourable to the occurrence of aftershocks. This clearly suggests a trigger mechanism at this period. Similar phase plots were obtained for this period with different pulse widths. Parts c) and d) of Figure 4 further substantiate our interpretation of the phase plots. The data window over which KORRECT was performed was shifted forward in time, just as in the case of spectral analysis. As seen in the left panel of Figure 4, the strength of tidal periodicities steadily declines with time; by the end of the 12[th] day after the mainshock, peaks at tidal periods vanish completely. Correspondingly, the phase variation curve of the KORRECT method for the semi-diurnal period exhibits a declining preference for the particular phases of the tidal cycle and by the end of the 12[th] day, it loses its systematic character. This behaviour is shown in Figure 10 for the 40 min pulse width.

From the combined results of the two independent analyses performed in this study, it is evident that the Earth tides had a strong influence on the occurrence times of aftershocks of the recent Tohoku-Oki earthquake. Two things can be said about the apparent weakening of the tidal-triggering effect with time. The first is that the steady decline in the number of aftershock makes it difficult to detect tidal periodicities. The second is that during the first few days after the mainshock, there is very high level of activity due to which the focal mechanisms of many aftershocks would be favourable to triggering by tidal stresses. A word of caution regarding the faithfulness of the results obtained by the KORRECT method would be well placed here. This method was originally developed to detect triggering of aftershocks by the free oscillations of the Earth. Free oscillations are normal modes, wherein the entire Earth oscillates in phase. In case of the Earth tide however, the induced variations are a function not only of time but also of location. This fact is overlooked in the application of KORRECT for detection of tidal triggering. However, we still place faith in the use of this method because aftershocks of an earthquake occur over a limited geographical region over which the spatial variation of tidal oscillations can be neglected in a zeroth order approximation. KORRECT results clearly corroborate the presence of unambiguous peaks in the spectral analysis.

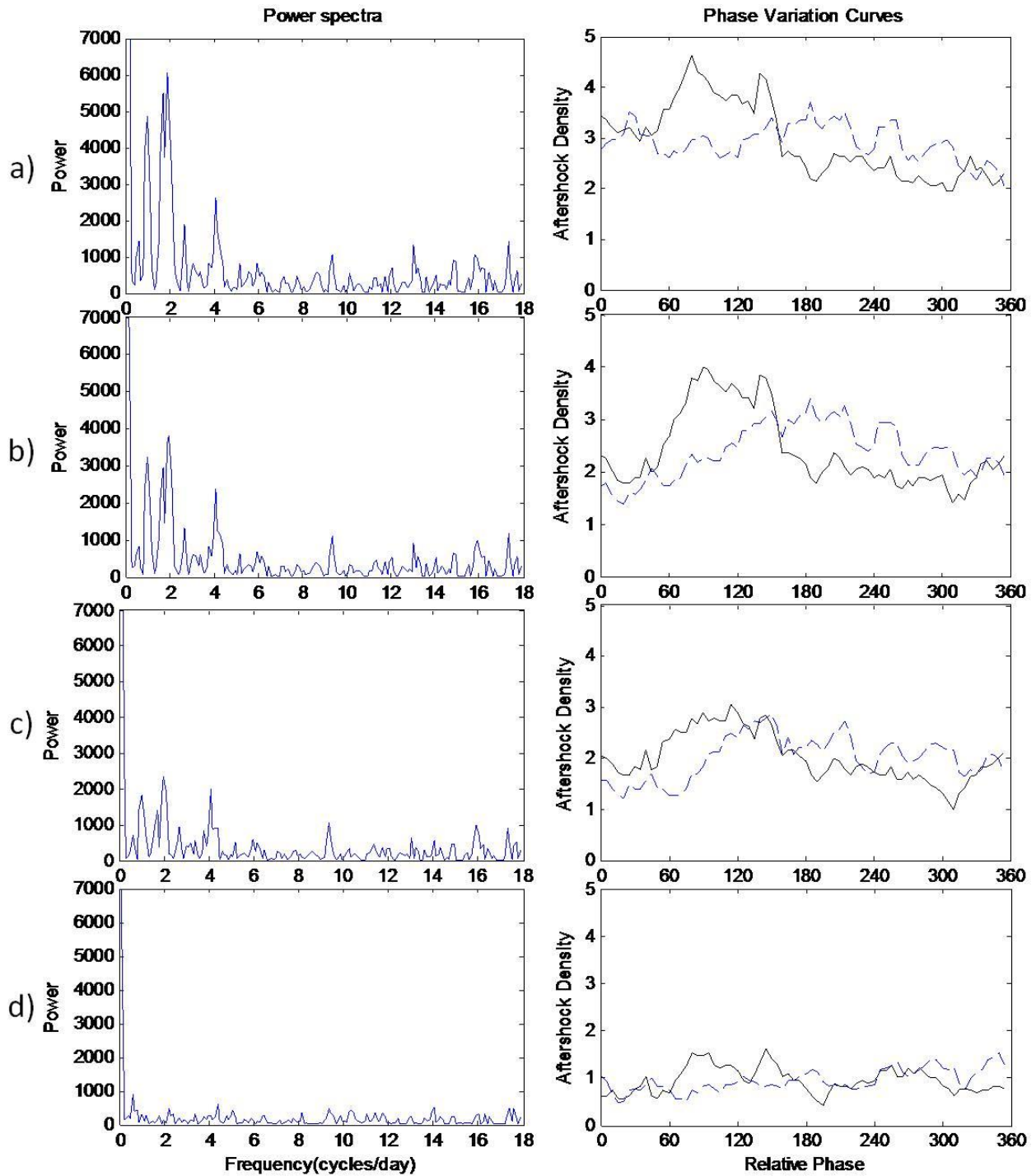

**Figure 4**: Comparison of frequency domain (left panel) and time-domain (right panel) results for a) 0 to 10 days, b) 0.5 to 10.5 days c) 1 to 11 days and d) 3 to 13 days after mainshock. Total length of data segment analyzed in both domains is constant at 240 hrs. Phase plots have been obtained for 12.63 hrs and also 10.5 hrs (dashed line) for comparison. Bin-width/pulse width is fixed at 40 min. In parts a) and b), note the trigger during the $60^0$-$150^0$ portion of the semi-diurnal tidal cycle. Also note that as the semi-diurnal spectral peak diminishes, its phase plot becomes increasingly unsystematic and similar to the phase plot for the 10.5 hr period, which is not associated with any physical phenomenon.

**Acknowledgement:** The freely available NEIC Earthquake Catalog data on the internet is gratefully acknowledged.

**References**

1. Scholz, C.H., The Mechanics of Earthquakes and Faulting. Cambridge Univ. Press, New York, 1990, 470p.
2. Vidale, J. E., D. C. Agnew, M. J. S. Johnston, and D. H. Oppenheimer, Absence of earthquake correlation with Earth tides: An indication of high preseismic fault stress rate, *J. Geophys. Res.*, 1998, **103**, 24,567–572.
3. Tanaka, S., Ohtake, M., Sato, H., Evidence for tidal triggering of earthquakes as revealed from statistical analysis of global data, *J. Geophys. Res.*, 2002, **107** (B10), 2211.doi:10.1029/2001JB001577.
4. Knott, C.G., On lunar periodicities in earthquake frequency, *Proc. Roy. Soc.*, 1896, **60**, 457-466.
5. Schuster, A., On lunar and solar periodicities of earthquakes, *Proc. R. Soc. London,* 1897, **61**, 455–465.
6. Emter, D., Tidal triggering of earthquakes and volcanic events. In *Tidal Phenomena, Lect. Notes Earth Sci.*, (ed. H. Wilhelm, W. Zurn, and H.-G. Wenzel), Springer-Verlag, New York, 1997, **66**, pp. 293 – 309.
7. Morgan,W. J., J. O. Stoner, and R. H. Dicke, Periodicity of earthquakes and the invariance of the gravitational constant, *J. Geophys. Res.*, 1961, **66**, 3831– 3843.
8. Tsuruoka, H., M. Ohtake, and H. Sato, Statistical test of the tidal triggering of earthquakes: Contribution of the ocean tide loading effect, *Geophys. J. Int.*, 1995, **122**, 183– 194.
9. Shlien, S., Earthquake-tide correlation, *Geophys. J. R. Astron. Soc.,* 1972, **28**, 27–34.
10. Shirley, J. H., Lunar and solar periodicities in large earthquakes: Southern California and the Alaska-Aleutian Islands seismic region, *Geophys. J.,* 1988, **92**, 403– 420.
11. Métivier, L., Viron, de O., Conrad, C., Renault, S., Diament, M., Patau, G., Evidence of earthquake triggering by the solid earth tides, *Earth Planet. Sci. Lett.,* 2009, **278**, 370-375.
12. Tanaka, S., Ohtake, M., Sato, H., Tidal triggering of earthquakes in Japan related to the regional tectonic stress, *Earth Planets Space*, 2004, **56**, 511-515.
13. Kamal and Mansinha, L., The Triggering of Aftershocks by the Free Oscillations of the earth, *Bull. Seismol. Soc. Am.,* 1996, **86**, 299-305.
14. Ryall, A., J. D. VanWormer, and A. E. Jones, Triggering of microearthquakes by earth tides, and other features of the Truckee, California, earthquake sequence of September, 1966, *Bull. Seismol. Soc. Am.,* 1968, **58**, 215–248.
15. Souriau, M., A. Souriau, and J. Gagnepain, Modeling and detecting interactions between Earth tides and earthquakes with application to an aftershock sequence in the Pyrenees, *Bull. Seismol. Soc. Am.,* 1982, **72**,165– 180.
16. Heaton, T. H., Tidal triggering of earthquakes, *Geophys. J. R. Astron. Soc.*, 1975, **43**, 307–326.